%% file: root.tex
\documentclass{llncs}

\usepackage{proof}
\usepackage{xspace}
\usepackage{amssymb,amsmath}

\newcommand{\ignore}[1]{}

\newcommand{\logic}{${\cal G}$\xspace}

\newcommand{\hsl}[1]{\hbox{\sl #1}}
\newcommand{\abs}[1]{\hsl {abs} \; #1}
\newcommand{\app}[2]{\hsl {app} \; #1 \; #2}
\newcommand{\step}[2]{\hsl {step} \; #1 \; #2}
\newcommand{\tabs}[2]{\hsl {abs} \; #1 \; #2}
\newcommand{\arr}[2]{\hsl {arr} \; #1 \; #2}
\newcommand{\of}[2]{\hsl {of} \; #1 \; #2}
\newcommand{\fresh}[2]{\hsl {fresh} \; #1 \; #2}
\newcommand{\name}[1]{\hsl {name} \; #1}

\newcommand{\tlam}[3]{\lambda #1 \! : \! #2 . #3}

\renewcommand{\vec}[1]{\overline{#1}}

\newcommand{\ie}{{\em i.e.}}
\newcommand{\eg}{{\em e.g.}}

\begin{document}

\title{The Abella Interactive Theorem Prover \\
(System Description)}
\author{Andrew Gacek}
\institute{Department of Computer Science and Engineering, University
  of Minnesota \\
200 Union Street SE, Minneapolis, MN 55455, USA
}
\maketitle

\thispagestyle{plain}

\input{introduction}
\input{foundations}
\input{architecture}
\input{implementation}
\input{examples}
\input{future-work}
\input{acknowledgements}

\bibliographystyle{plain}
\bibliography{root}

\end{document}

%% file: introduction.tex
\vspace{-0.3cm}
\section{Introduction}
\vspace{-0.1cm}

Abella \cite{abella-website} is an interactive system for
reasoning about aspects of object languages that have been formally
presented through recursive rules based on syntactic structure.
Abella utilizes a two-level logic approach to specification and
reasoning. One 
level is defined by a specification logic which supports a transparent
encoding of structural semantics rules and also enables
their execution. The second level, called the reasoning logic, embeds
the specification logic and allows the development of proofs of
properties about specifications. An important characteristic of both
logics is that they exploit the $\lambda$-tree syntax approach to
treating binding in object languages. Amongst other things,
Abella has been used to prove normalizability properties of the
$\lambda$-calculus, cut admissibility for a sequent calculus and type
uniqueness and subject reduction properties. This paper 
discusses the logical foundations of Abella, outlines the style of
theorem proving that it supports and finally describes some of its
recent applications.


%% file: foundations.tex
\vspace{-0.1cm}
\section{The Logic Underlying Abella}
\label{sec:foundations}
\vspace{-0.1cm}

Abella is based on \logic, an intuitionistic, predicative,
higher-order logic with fixed-point definitions for atomic
predicates and with natural number induction \cite{gacek08lics}.

\vspace{-0.2cm}
\paragraph{Representing binding.}

\logic uses the {\em $\lambda$-tree syntax} approach to representing
syntactic structures \cite{miller00cl}, which allows object level
binding to be represented using meta-level abstraction. Thus common
notions related to binding such as $\alpha$-equivalence and
capture-avoiding substitution are built into the logic, and the
encodings of object
languages do not need to implement such features.

To reason over $\lambda$-tree syntax, \logic uses the $\nabla$
quantifier which represents a notion of generic judgment
\cite{miller05tocl}. A formula $\nabla x. F$ is true if $F$ is true
for each $x$ in a generic way, \ie, when nothing is assumed about any $x$.
This is a stronger statement than $\forall x. F$ which says that $F$
is true for all values for $x$ but allows this to be shown in
different ways for different values.

For the logic \logic, we assume the following two properties of
$\nabla$:
\vspace{-0.1cm}
\begin{equation*}
\nabla x. \nabla y. F\; x\; y \equiv \nabla y. \nabla x. F\; x\; y
\hspace{0.5cm} \nabla x. F \equiv F \mbox{ if $x$ not free in $F$}
\vspace{-0.1cm}
\end{equation*}
A natural proof-theoretic treatment for this quantifier is to use {\em
  nominal constants} to instantiate $\nabla$-bound variables
  \cite{tiu06lfmtp}. Specifically, the 
proof rules for $\nabla$ are
\begin{align*}
\infer[\nabla\mathcal{L}]{\Gamma, \nabla x. B \vdash C}{\Gamma, B[a/x]
  \vdash C} &&
\infer[\nabla\mathcal{R}]{\Gamma \vdash \nabla x. C}{\Gamma \vdash
  C[a/x]} 
\end{align*}
where $a$ is a nominal constant which does not appear in the formula
underneath the $\nabla$ quantifier.
Due to the equivalence of permuting $\nabla$ quantifiers,
nominal constants must be treated as permutable, which is captured by
the initial rule.
\vspace{-0.1cm}
\begin{equation*}
\infer[id_\pi]{\Gamma, B \vdash B'}{\pi.B = B'}
\vspace{-0.1cm}
\end{equation*}
Here $\pi$ is a permutation of nominal constants.

\vspace{-0.2cm}
\paragraph{Definitions.}

The logic \logic supports fixed-point definitions of atomic
predicates. These definitions are specified as clauses of the form
$\forall \vec{x} . (\nabla \vec{z}. H) \triangleq B$ where the head
$H$ is an atomic predicate. This notion of definition is extended from
previous notions (\eg, see \cite{miller05tocl}) by admitting the
$\nabla$-quantifier in the head. Roughly, when such a definition is
used, in ways to be explained soon, these $\nabla$-quantified
variables become instantiated with nominal constants from the term on
which the definition is used. The instantiations for the universal
variables $\vec{x}$ may contain any nominal constants not assigned to
the variables $\vec{z}$. Thus $\nabla$ quantification in the head of a
definition allows us to restrict certain pieces of syntax to be
nominal constants and to state dependency information for those
nominal constants.

Two examples hint at the expressiveness of our extended form of
definitions. First, we can define a predicate $\name E$ which holds
only when $E$ is a nominal constant. Second, we can define a
predicate $\fresh X E$ which holds only when $X$ is a nominal
constant which does not occur in $E$.
\vspace{-0.1cm}
\begin{equation*}
(\nabla x. \name x) \triangleq \top
\hspace{1cm}
\forall E. (\nabla x. \fresh x E) \triangleq \top
\vspace{-0.1cm}
\end{equation*}
Note that the order of quantification in {\sl fresh} enforces the
freshness condition.

Definitions can be used in both a positive and negative fashion.
Positively, definitions are used to derive an atomic judgment, \ie, to
show a predicate holds on particular values. This use corresponds to
unfolding a definition and is similar to back-chaining. Negatively, an
atomic judgment can be decomposed in a case analysis-like way based on
a closed-world reading of definitions. In this case, the atomic
judgment is unified with the head of each definitional clause, where
eigenvariables are treated as instantiatable. Also, both the positive and
negative uses of definitions consider permutations of nominal
constants in order to allow the $\nabla$-bound variables $\vec{z}$ to
range over any nominal constants.
A precise presentation of these rules, which is provided in
Gacek {\it et al.}~\cite{gacek08lics}, essentially amounts to introduction rules for
atomic judgments on the right and left sides of sequents in a sequent
calculus based presentation of the logic.

\vspace{-0.2cm}
\paragraph{Induction.}

\logic supports induction over natural numbers. By augmenting the
predicates being defined with a natural number argument, this
induction can serve as a method of proof based on the length of a
bottom-up evaluation of a definition.


%% file: architecture.tex
\vspace{-0.1cm}
\section{The Structure of Abella}
\label{sec:architecture}
\vspace{-0.1cm}

The architecture of Abella has two distinguishing characteristics. 
First, Abella is oriented towards the use of a specific (executable)
specification logic whose proof-theoretic structure is encoded via
definitions in \logic. 
Second, Abella provides tactics for proof construction that embody
special knowledge of the specification logic. We discuss these aspects
and their impact in more detail below. 

\vspace{-0.1cm}
\subsection{Specification Logic}
\vspace{-0.1cm}

It is possible to encode object language descriptions directly in
definitions in \logic, but there are two disadvantages to doing so:
the resulting definitions may not be 
executable and there are common patterns in specifications with
$\lambda$-tree syntax which we would like to take advantage of. We
address these issues by selecting a specification logic which has the
features that the \logic lacks, and embedding the
evaluation rules of this specification logic instead into
\logic. Object languages are then encoded through descriptions in the
specification logic \cite{mcdowell02tocl}.

The specification logic of Abella is second-order hereditary Harrop
formulas \cite{miller91apal} with support for $\lambda$-tree syntax.
This allows a transparent encoding of structural operational semantics
rules which operate on objects with binding. For example, consider the
simply-typed $\lambda$-calculus where types are either a base type
$i$ or arrow types constructed with {\sl arr}. Terms are encoded
with the constructors {\sl app} and {\sl abs}. The constructor {\sl
  abs} takes two arguments: the type of the variable being
abstracted and the body of the function. Rather than having a
constructor for variables, the body argument to {\sl abs} is an
abstraction in our specification logic, thus object level binding is
represented by the specification logic binding. For example, the term
$(\tlam f {i\to i} (\tlam x i (f\; x)))$ is encoded as
\vspace{-0.2cm}
\begin{equation*}
\tabs {(\arr i i)} (\lambda f. \tabs i (\lambda x. \app f
x)).
\vspace{-0.2cm}
\end{equation*}
In the latter term, $\lambda$ denotes an abstraction in the specification
logic. Given this representation, the typing judgment $\of m t$ is
defined in Figure~\ref{fig:typing}. Note that these rules do not
maintain an explicit context for typing assumptions, instead using a
hypothetical judgment to represent assumptions. Also, there is no
side-condition in the rule for typing abstractions to ensure the
variable $x$ does not yet occur in the typing context, since instead
of using a particular $x$ for recording a typing assumption, we
quantify over all $x$.

\begin{figure}[t]
\vspace{-0.5cm}
\begin{center}
\begin{tabular}{c}
$\forall m, n, a, b[\of m (\arr a b) \land
    \of n a \; \supset \; \of{(\app m n)} b]$\\[6pt] 
$\forall r, a, b[\forall x[\of x a  \supset 
    \of{(r \; x)}{b}] \supset \of{(\tabs a r)}{(\arr a b)}]$
\end{tabular}
\end{center}
\vspace{-0.5cm}
\caption{Second-order hereditary Harrop formulas for typing}
\label{fig:typing}
\vspace{-0.5cm}
\end{figure}

Our specification of typing assignment is executable. More generally,
the Abella specification logic is a subset of the language
$\lambda$Prolog \cite{nadathur88iclp} which can be compiled and
executed efficiently \cite{nadathur99cade}. This enables the animation
of specifications, which is convenient for assessing specifications
before attempting to prove properties over them. This also allows
specifications to be used as testing oracles when developing full
implementations.

The evaluation rules of our specification logic can be encoded as a
definition in \logic. A particular specification is then encoded in a
separate definition which is used by the definition of evaluation in
order to realize back-chaining over specification clauses. Reasoning over a
specification is realized by reasoning over its evaluation via the
definition of the specification logic. Abella takes this 
further and is customized towards the specification logic. For
example, the context of hypothetical judgments in our specification
logic admits weakening, contraction, and permutation, all of which are provable in
\logic. Abella automatically uses this meta-level property of the
specification logic during reasoning. Details on the benefits of
this approach to reasoning are available in Gacek {\it et al.}
\cite{gacek08lfmtp}.

\vspace{-0.1cm}
\subsection{Tactics}
\vspace{-0.1cm}

The user constructs proofs in Abella by applying tactics which
correspond to high-level reasoning steps. The collection of tactics
can be grouped into those that generically orchestrate the
rules of \logic and those that correspond to meta-properties of the
specification logic. We discuss these classes in more detail below.

\vspace{-0.2cm}
\paragraph{Generic tactics.}

The majority of tactics in Abella correspond directly to inference
rules in \logic. The most common tactics from this group are the ones
which perform induction, introduce variables and hypotheses, conduct
case analysis, apply lemmas, and build results from hypotheses. In the
examples suite distributed with Abella, these five tactics make up
more than 90\% of all tactic usages. The remaining generic
tactics are for tasks such as splitting a goal of the form $G_1 \land
G_2$ into two separate goals for $G_1$ and $G_2$, or for instantiating
the quantifier in a goal of the form $\exists x . G$.

\vspace{-0.2cm}
\paragraph{Specification logic tactics.}

Since our specification logic is encoded in \logic, we can formally
prove meta-level properties for it. Once such properties are proved,
their use in proofs can be built into tactics. Two important
properties that Abella uses in this way are instantiation and cut
admissibility. In
detail, negative uses of the 
specification logic $\forall$ quantifier are represented in \logic as
nominal constants (\ie, the $\nabla$ quantifier), and the instantiation tactic allows such nominal
constants to be instantiated with specific terms. The cut tactic
allows hypothetical judgments to be relieved by showing that they are
themselves provable.



%% file: implementation.tex
\vspace{-0.1cm}
\section{Implementation}
\label{sec:implementation}
\vspace{-0.1cm}

Abella is implemented in OCaml. The most sophisticated component of
this implementation is higher-order unification which is a
fundamental part of the logic \logic. It underlies how case analysis
is performed, and in the implementation, unification is used to decide
when tactics apply and to determine their result. Thus an efficient
implementation of higher-order unification is central to an efficient
prover. For this, Abella uses the the higher-order pattern unification
package of Nadathur and Linnell \cite{nadathur05iclp}. We have also
extended this package to deal with the particular features and
consequences of reasoning in \logic.

\vspace{-0.2cm}
\paragraph{Treatment of nominal constants.}

As their name suggests, nominal constants can be treated very
similarly to constants for most of the unification algorithm, but
there are two key differences. First, while traditional constants can
appear in the instantiation of variables, nominal constants
cannot appear in the instantiation of variables. Thus dependency
information on nominal constants is tracked via explicit raising of
variables. Second, nominal constants can be permuted when determining
unifiability. However, even in our most sophisticated examples
the number of nominal constants appearing
at the same time has been at most two. Thus, naive approaches to
handling permutability of nominal constants have sufficed and
there has been little need to develop sophisticated algorithms.

\vspace{-0.2cm}
\paragraph{Simple extensions.}

The treatment of case analysis via unification for eigenvariables
creates unification problems which fall outside of the higher-order
pattern unification fragment, yet still have most general unifiers.
For example, consider the clause for $\beta$-contraction in the
$\lambda$-calculus:
\vspace{-0.1cm}
\begin{equation*}
\step {(\app {(\abs R)} M)} (R\; M).
\vspace{-0.1cm}
\end{equation*}
Case analysis on a
hypotheses of the form $\step A B$ will result in the attempt to solve
the unification problem $B = R\; M$ where $B$, $R$, and $M$ are all
instantiatable. This is outside of the higher-order pattern unification
fragment since $R$ is applied to an instantiatable variable, but there
is a clear most general unifier. When nominal constants are present,
this situation is slightly more complicated with unification problems
such as $B\; x = R\; M\; x$ or $B\; x = R\; (M\; x)$, where $x$ is a
nominal constant. The result is the same, however, that a most general
unifier exists and is easy to find.



%% file: examples.tex
\vspace{-0.1cm}
\section{Examples}
\label{sec:examples}
\vspace{-0.1cm}

This section briefly describes sample reasoning tasks we
have conducted in Abella. The detailed proofs are available in
the distribution of Abella \cite{abella-website}.

\vspace{-0.2cm}
\paragraph{Results from the $\lambda$-calculus.}
Over untyped $\lambda$-terms, we have shown the equivalence of
big-step and small-step evaluation, preservation of typing for both
forms of evaluation, and determinacy for both forms of evaluation.
We have shown that the $\lambda$-terms can be disjointly
partitioned into normal and non-normal forms. Over simply-typed
$\lambda$-terms, we have shown that typing assignments are unique.

\vspace{-0.2cm}
\paragraph{Cut admissibility.}
We have shown that the cut rule is admissible for a sequent calculus
with implication and conjunction. The representation of sequents in
our specification logic used hypothetical judgments to represent
hypotheses in the sequent. This allowed the cut admissibility proof to
take advantage of Abella's built-in treatment of meta-properties of
the specification logic.

\vspace{-0.2cm}
\paragraph{The POPLmark challenge.}
The POPLmark challenge \cite{aydemir05tphols} is a selection of
problems which highlight the traditional difficulties in reasoning
over systems which manipulate objects with binding. The particular
tasks of the challenge involve reasoning about evaluation, typing, and
subtyping for $F_{<:}$, a $\lambda$-calculus with bounded subtype
polymorphism. We have solved parts 1a and 2a of this challenge using
Abella, which represent the fundamental reasoning tasks involving
objects with binding.

\vspace{-0.2cm}
\paragraph{Proving normalizability {\`a} la Tait.}
We have shown that all closed terms in the call-by-value, simply-typed
$\lambda$-calculus are
normalizable using the logical relations argument in the style
of Tait \cite{tait67jsl}. Fundamental in this proof was
the encoding of arbitrary cascading substitutions which allows one to
consider all closed instantiations for an open $\lambda$-term.
Encoding and reasoning over this form of substitution makes essential
use of the extended form of definitions in \logic.


%% file: future-work.tex
\vspace{-0.1cm}
\section{Future and Related Work}
\label{sec:future-work}

\vspace{-0.2cm}
\paragraph{Induction and coinduction}

The logic \logic currently supports induction on natural numbers.
Similar logics have been extended to support structural induction and
coinduction on definitions \cite{tiu04phd}. Already, the
implementation of Abella has support for these features. A paper which
describes the extended logic supporting these features is in
preparation.

\vspace{-0.2cm}
\paragraph{User programmability.}

Tactics-based theorem provers often support {\em tacticals} which
allow users to compose tactics in useful ways. Some systems even go
beyond this and offer a full programming language for creating custom
tactics. We would like to extend Abella with such features.

\vspace{-0.2cm}
\paragraph{Proof search.}

Many proofs in Abella follow a straightforward pattern of essentially
induction, case analysis, and building from hypotheses. We would like
to extend Abella to perform these types of proofs automatically.
Recent results on focusing in similar logics may offer some insight
into a disciplined approach to automated proof search \cite{baelde07lpar}.

\vspace{-0.2cm}
\paragraph{Related work.}

A closely related system is Twelf \cite{pfenning99cade} which is based
on a dependently typed $\lambda$-calculus for specification.
Controlling for
dependent types, the most significant difference is that our
meta-logic is significantly richer than the one in Twelf. Also related
is the Nominal package \cite{urban05cade} for Isabelle/HOL which
allows for reasoning over $\alpha$-equivalence classes. This approach
leverages on existing theorem proving work, but does not address the
full problem of reasoning with binding. In particular, all work
related to substitution is left to the user. A more detailed
comparison with these works is available in Gacek {\it et al.}
\cite{gacek08lfmtp}.



%% file: acknowledgements.tex
\vspace{-0.1cm}
\section*{Acknowledgements}
\vspace{-0.1cm}
I am grateful to David Baelde, Dale Miller, Gopalan Nadathur, Randy
\mbox{Pollack}, and
Alwen Tiu for their input and feedback on the development of Abella. 
Anonymous reviewers provided helpful comments on an 
earlier version of this paper.
This work has been supported by 
the NSF Grant CCR-0429572 and by a grant
from Boston Scientific. Opinions, findings, and conclusions or
recommendations expressed in this work are those of the authors and do
not necessarily reflect the views of the National Science Foundation.

